\documentclass[fleqn,preprint,showpacs,preprintnumbers,amsmath,amssymb]{revtex4}
\usepackage{graphicx}
\usepackage{dcolumn}
\usepackage{bm}

\begin{document}

\title{Characteristics of Quantum Magnetosonic-Wave Dispersion}
\author{M. Akbari-Moghanjoughi}
\affiliation{Azarbaijan University of
Tarbiat Moallem, Faculty of Sciences, Department of Physics,
51745-406, Tabriz, Iran}

\date{\today}
\begin{abstract}
Using the quantum magnetohydrodynamics (QMHD) model, linear dispersion of magnetosonic waves are studied in a quasineutral quantum electron-ion plasma in two distinct regimes of nonrelativistic and relativistic degeneracies considering also the plasma composition effect. It is shown that the paramagnetic spin effects of the degenerated electrons plays a key role in dynamics of magnetosonic waves. The linear wave-speed is found to have minimum value at some degeneracy parameter in such plasmas. This is due to delicate interplay between relativistic degeneracy and the Pauli spin-magnetization. It is also revealed that the plasma composition has significant effect on the linear dispersion in the relativistic degeneracy limit contrary to that of nonrelativistic case and Zeeman energy has significant effect in nonrelativistic degeneracy regime unlike that of relativistic one in the linear perturbation-limit. Current findings can have important applications in both inertial plasma confinement and astrophysical degenerate plasmas.
\end{abstract}

\maketitle

\section{Introduction}

A quantum plasma is usually referred to as a degenerate dense and cold ionized matter. Despite the high temperatures observed for astrophysical compact objects, it has been shown \cite{chandra1} that these objects can be regarded as fully degenerate cold quantum plasmas. This is because the temperature of plasma species in these objects stay below a well defined Fermi temperature of order $10^4K$. Vast majority of the recent studies on quantum plasmas \cite{bonitz, haas1, haas2, Markowich, Marklund1, Brodin1, Marklund2, Brodin2, akbari1, akbari2} have been inspired by the pioneering works of Bohm, Pines and Levine et.al \cite{bohm, pines, levine} and the earlier works of Chandrasekhar. Distinction between an ordinary and quantum plasma (also known as the zero-temperature Fermi-plasma) is that, in a quantum plasma contrary to the ordinary one, the inter-fermion distances are much lower than the characteristic de Broglie thermal wavelength, $h/(2\pi m k_B T)^{1/2}$ so that the particle waves overlap causing a new type collective phenomenon through the electron degeneracy and tunneling effects. These properties which are encountered only in quantum plasmas, for instance in ordinary metals and compact astrophysical objects, lead to distinct features of wave dynamics when the electron number-density or the ambient magnetic field strength or even the system dimensionality and the degrees of freedom of fermions is varied \cite{akbari3, akbari4, akbari5}. The electron degeneracy is a direct consequence of the Pauli exclusion principle and effectively rules the thermodynamical properties of a dense matter \cite{landau}.

Most of current investigation on quantum plasmas are motivated due to the applicability of the quantum magnetohydrodynamics (QMHD) to semiconductors \cite{gardner}. Quantum optical and electronic transport phenomena are also known as main ingredients in the study of metallic and semiconductor nano-structured materials such as, nano-particles, quantum-wells, quantum-wires and quantum-dots \cite{haug}. Therefore, the subject of dense cold ionized matter under strong magnetic field can promise wide applications in both laboratory as well as naturally occurring plasmas such as astrophysical compact objects \cite{manfredi, shukla}. One of the differences of the QMHD model from those of classical counterparts is the inclusion Bohm tunneling effect. It has been shown that the increase in the electron number density enhances this feature significantly altering the dynamics of linear and nonlinear wave propagations \cite{sabry}. More recently, Marlkund and Brodin have extended QMHD model and incorporated the electron spin-1/2 effects \cite{Marklund3} by introducing a generalized term for the quantum force a feature which is expected to be dominant in superdense plasmas such as Fermi-Dirac. It has been estimated that relativistic
electrons in a white dwarf can generate magnetic fields of the order of $10^7G$ and higher \cite{lee}, and that in a neutron star the self-field may be of order
$10^{13}-10^{14}G$ \cite{can}. On the other hand, it has been shown that, in a perfectly conductive quantum plasma, the spin magnetization term introduced in quantum magnetohydrodynamics (QMHD) equations acts as a negative pressure-like entity significantly affecting the dynamics of spin-induced nonlinear magnetosonic excitations \cite{Marklund4}.

In 1935 Chandrasekhar \cite{chandra2}, in a pioneering work, proved that the gigantic inward gravitational pressure can lead to the new state of relativistic degeneracy for electrons and may cause the ultimate collapse of the star in extreme degeneracy cases \cite{chandra3}. The sudden collapse of the white dwarfs was shown to be due to softening of the degeneracy pressure of electrons relativistic motion. This effect has been shown to alter the whole thermodynamical properties of the star \cite{kothary}. Current investigation is an attempt to trace possible differences in linear magnetosonic wave dynamics caused due to change in the plasma degeneracy state such as nonrelativistic and relativistic ones and explore the composition effects on linear dispersion. The presentation of the article is as follows. The QMHD plasma model including the spin contribution is introduced in Sec. \ref{equations}. The linear dispersion of spin-induced magnetoacoustic waves are derived in Sec. \ref{extreme} for different plasma degeneracy regimes and the numerical results are presented in Sec. \ref{discussion}.

\section{QMHD Model Including Spin Effects}\label{equations}

Let us consider a collisionless dense degenerate quasineutral electron-ion perfectly conducting plasma with electrons and ions immersed in a spatially varying magnetic field in a fixed direction. We further assume that the magnitude of the external field is very high and the dynamics of wave propagation is strongly affected by spin contribution from fermions. In the center of mass frame of plasma the continuity relation reads as
\begin{equation}\label{cont}
\frac{{\partial \rho }}{{\partial t}} + \nabla  \cdot (\rho {\bf{V}}) = 0,
\end{equation}
where, $\rho=m_e n_e + m_i n_i\simeq m_i n_i$ and $\bm{V}=(n_e m_e \bm{V_e} + n_i m_i \bm{V_i})/\rho$ are the plasma center of mass density and velocity. The momentum equation, taking into account the quantum diffraction \cite{haas1} and spin pressure \cite{Marklund3} effects, in the center of mass frame takes the form \cite{Marklund4, misra}
\begin{equation}\label{mom}
\frac{{\partial {\bf{V}}}}{{\partial t}} + \left( {{\bf{V}} \cdot \nabla } \right){\bf{V}} = {\rho ^{ - 1}}\left( {{\bf{J}} \times {\bf{B}} - \nabla P + {{\bf{F}}_Q}} \right),
\end{equation}
where, the quantum force $\bm{F}_Q={\bm{F}}_B+{\bm{F}}_S$ is a collective contribution of quantum Bohm-force, ${\bm{F}}_B$, and the so-called spin-force, ${\bm{F}}_S$, namely,
\begin{equation}\label{QF}
{{\bf{F}}_Q} = \frac{{{\rho_c\hbar ^2}}}{{2{m_e}{m_i}}}\nabla \frac{{\Delta \sqrt {{\rho _c}} }}{{\sqrt {{\rho _c}} }} + {{M}} \nabla {{B}}.
\end{equation}
The quantities $\bf{J}$, $\bf{B}$ and $P$ are the magnetization current, magnetic field and scalar total center of mass plasma pressure, respectively. The magnetization current, $\bf{J}$, is related to the spin-magnetization (Pauli-magnetization) per unit volume ${{M}_P} = \left( {3\mu _B^2{n_e}/2{k_B}{T_{Fe}}} \right){{B}}$ ($T_{Fe}$ being the electron Fermi-temperature) through; ${\bf{J}} = \mu _0^{ - 1}\nabla  \times ({\bf{B}} - {\mu _0}{\bf{M}})$, where, $\mu_B=e\hbar/2m_e c$ is the Bohr magneton and $\hbar$ is the normalized Plank constant. It should be noted that, in the case of fully degenerate quantum plasma the Pauli paramagnetism dominates the Langevin-Type one, hence, the Fermi-temperature, $T_{Fe}$, replaces the thermal electron temperature, $T_e$ in magnetic pressure term. In a completely degenerate plasma, due to the Pauli exclusion principal, only a small fraction ($T_e/T_{Fe}$), i.e. unpaired free-electrons, with energies nearly equal to the Fermi-energy contribute to the collective spin effects. This effect gives rise to the well-known Pauli paramagnetic law of the form $m_P = 3N\mu _B^2B/(2{E_{Fe}})$ \cite{landau} instead of the classical law of $m_c = N{\mu _B}\tanh ({\mu _B}B/{k_B}{T_e})$ frequently used for dense plasmas in literature. On the other hand, the Landau demagnetization factor of $1/3$ of this value should be subtracted, which results in total sum of $M = M_P + M_L = n_e\mu _B^2B/{E_{Fe}}$. The susceptibility of weakly interacting relativistic Fermi-gas in a weak magnetic field ($\mu_B B_0 \ll m_e c^2$) has been evaluated in Ref. \cite{men}, which is as follows
\begin{equation}\label{QP}
{\chi _p} = \left( {\frac{{3n\mu _B^2}}{{{m_e}{c^2}}}} \right){\left( {\frac{{{m_e}c}}{{{P_{Fe}}}}} \right)^2}\sqrt {1 + {{\left( {\frac{{{P_{Fe}}}}{{{m_e}c}}} \right)}^2}} \left[ {1 + \frac{{3\gamma n}}{{2{m_e}{c^2}}}{{\left( {\frac{{{m_e}c}}{{{P_{Fe}}}}} \right)}^2}\sqrt {1 + {{\left( {\frac{{{P_{Fe}}}}{{{m_e}c}}} \right)}^2}} } \right],
\end{equation}
where, $\gamma$ and $P_{Fe}$ are the s-wave interaction parameter and the Fermi relativistic momentum, respectively. It can be shown that in the nonrelativistic limit with $\gamma=0$ the susceptibility given in Eq. (\ref{QP}) reduces to the well-known Pauli paramagnetic susceptibility. In order to complete the set of QMHD fluid equations, we use the generalized Faraday law without the Hall term in the form
\begin{equation}\label{FL}
\frac{{\partial {\bf{B}}}}{{\partial t}} = \nabla  \times \left\{ {{\bf{V}} \times {\bf{B}} - \eta {\bf{J}} - \frac{{{\bf{J}} \times {\bf{B}} + {{\bf{F}}_Q}}}{{e{n_e}}} - \frac{{{m_e}}}{{{e^2}{\mu _0}}}\left[ {\frac{\partial }{{\partial t}} - \left( {\frac{{\nabla  \times {\bf{B}}}}{{e{\mu _0}{n_e}}}} \right) \cdot \nabla } \right]\frac{{\nabla  \times {\bf{B}}}}{{{n_e}}}} \right\}.
\end{equation}
where, $\eta$ is the plasma resistivity. Moreover, with $\rho  = \rho (x,t)$ and $V = V(x,t)$, and assuming the propagation of the magnetoacoustic wave to be in $x$ direction perpendicular to the magnetic field $B(x,t)$ along the $z$-axis, we may write simplified one-dimensional version of the complete QMHD equations as \cite{Marklund4}
\begin{equation}\label{dimensional}
\begin{array}{l}
\frac{{\partial {\rho}}}{{\partial t}} + \frac{{\partial {\rho}{V}}}{{\partial x}} = 0,\hspace{3mm}{\rho} = {\rho}(x,t),\hspace{3mm}{V} = {V}(x,t),\hspace{3mm}B = B(x,t),\\ \frac{{\partial {V}}}{{\partial t}} + {V}\frac{{\partial {V}}}{{\partial x}} + \frac{B}{{{\mu _0}{\rho}}}\frac{{\partial B}}{{\partial x}} + \frac{1}{{{\rho}}}\frac{{\partial {P}}}{{\partial x}} - \frac{{{\hbar ^2}}}{{2{m_e}{m_i}}}\frac{\partial }{{\partial x}}\left( {\frac{1}{{{{\sqrt \rho  }}}}\frac{{{\partial ^2}{{\sqrt \rho  }}}}{{\partial {x^2}}}} \right) \\ - \frac{{3\mu _B^2}}{{{m_e}{m_i}{\rho}{c^2}}}\frac{\partial }{{\partial x}}\left\{ {{\rho}{B^2}{{\left( {\frac{{{m_e}c}}{{{P_{Fe}}}}} \right)}^2}\sqrt {1 + {{\left( {\frac{{{P_{Fe}}}}{{{m_e}c}}} \right)}^2}} } \right\} = 0, \\ \frac{{\partial B}}{{\partial t}} + \frac{{\partial B{V}}}{{\partial x}} - \frac{\eta }{{{\mu _0}}}\frac{{{\partial ^2}B}}{{\partial {x^2}}} = 0 \\
\end{array}
\end{equation}
On the other hand, in a relativistically degenerate zero-temperature Fermi-Dirac plasma one can ignore the ion-pressure compared to that of the dominant electron degeneracy pressure \cite{akbari6} which can be expressed in the following general form \cite{chandra1}
\begin{equation}\label{p}
{P_e} = \frac{{\pi m_e^4{c^5}}}{{3{h^3}}}\left\{ {R\left( {2{R^2} - 3} \right)\sqrt {1 + {R^2}}  + 3\ln \left[ {R + \sqrt {1 + {R^2}} } \right]} \right\}{\rm{ }},
\end{equation}
in which $R=(P_{Fe})/(m_e c)=(\rho/\rho_{c})^{1/3}$ (${\rho_c}\simeq 1.97 \times 10^{6} gr/cm^{3}$) is the relativity parameter with $P_{Fe}$ being the electron Fermi relativistic momentum. The relativistic degeneracy due to the Pauli exclusion principle increases with decrease in inter-fermion distances in degenerated plasmas. It is noted that, although the relativistic effects arise due to increase in fermion number-density in a dense degenerate plasma, however, unlike for the low-pressure relativistic plasmas the degeneracy pressure in the fermion momentum fluid equation usually dominates the relativistic (electron/ion) dynamic effects in super-dense plasma state and the relativistic electron/ion mass effects may be ignored. It has been shown \cite{chandra2} that for a zero-temperature dense degenerate Fermi-gas such as a white-dwarf with a mass-density, $\rho$, the degeneracy pressure turns from $P_e\propto \rho^{5/3}$ (with polytropic index $3$) dependence for normal degeneracy in the limit $R \ll 1$ to $P_e\propto \rho^{4/3}$ (with polytropic index $2/3$) dependence for relativistic degeneracy case in the limit $R \gg 1$. On the other hand, the relativistic degeneracy starts at mass density of about $4.19\times 10^6 (gr/cm^3)$ of the order in the core of a $0.3M_\odot$ white dwarf, which corresponds to a Fermi-momentum $P_{Fe}\sim 1.29 m_e c$ (equivalent to the relativistic degeneracy parameter value of $R\sim 1.29$) or the threshold velocity of $u_{Fe}\sim 0.63c$ (the Fermi relativistic factor $\gamma_{Fe}\sim 1.287$). This may justify our nonrelativistic treatment of electron mass effects in above arguments.

The problem of spin-induced magnetosonic has been thoroughly investigated by Masood et al. \cite{masood} for the case of the relativistic degeneracy, however, for completeness in current work we try to extend the previous study to the case of quantum plasma with the Coulomb interaction effect. In a dense quantum plasma the coupling parameter $\Gamma$ representing the potential to kinetic energy ratio may exceed unity leading to strong coupling due to other many-body effects such as Thomas-Fermi screening, Coulomb interactions, electron-exchange and ion-correlation. Salpeter \cite{salpeter} has examined these effects in the relativistically degenerate plasmas and concluded that the first-order correction to the degeneracy pressure can be that of Coulomb interactions and other effects may be negligible. Following the Salpeter expression for classical Coulomb plus Thomas-Fermi corrections, we may write
\begin{equation}\label{ctf}
{P_{C + TF}} =  - \frac{{8{\pi ^3}{m_e^4}{c^5}}}{{{h^3}}}\left[ {\frac{{\alpha {Z^{2/3}}}}{{10{\pi ^2}}}{{\left( {\frac{4}{{9\pi }}} \right)}^{1/3}}{R^4} + \frac{{162}}{{175}}\frac{{{{(\alpha {Z^{2/3}})}^2}}}{{9{\pi ^2}}}{{\left( {\frac{4}{{9\pi }}} \right)}^{2/3}}\frac{{{R^5}}}{{\sqrt {1 + {R^2}} }}} \right].
\end{equation}
where, $\alpha=e^2/\hbar c\simeq 1/137$ and $Z$ are the fine-structure constant and the atomic number, respectively. The first term corresponds to the Coulomb negative pressure and the second term (proportional to $\alpha^2$) is a relatively smaller contribution due to Thomas-Fermi screening, which will be neglected in current scheme.

On the other hand, the plasma resistivity, $\eta$, of a completely degenerate Fermi-Dirac plasma is supposed to be negligible, since, the electron-ion collisions are lowered by Pauli-blocking mechanism, i.e., $\eta\sim 0$. By using Eq. (\ref{dimensional}) this lead to the simple relation, $B/B_0=\rho/\rho_{0}=\bar{\rho}$ (the so-called frozen-in field condition). The new quantities $B_0$, $\rho_{0}$ and $\bar{\rho}$ denote the equilibrium values of magnetic field intensity, plasma mass-density and the normalized mass density, respectively. Therefore, the complete set of dimensionless equations governing the dynamics of magnetosonic waves reduce to
\begin{equation}\label{diff}
\begin{array}{l}
\frac{{\partial \rho }}{{\partial t}} + \frac{{\partial \rho V}}{{\partial x}} = 0, \\
\frac{{\partial V}}{{\partial t}} + V\frac{{\partial V}}{{\partial x}} =  \\
- \frac{{{\epsilon^2}}}{{{H^2}}}\frac{{\partial \rho }}{{\partial x}} - \frac{\partial }{{\partial x}}\left[ {\sqrt {1 + R_0^2{\rho ^{2/3}}}  - \beta {R_0}{\rho ^{1/3}}} \right] + {H^2}\frac{\partial }{{\partial x}}\left( {\frac{1}{{\sqrt \rho  }}\frac{{{\partial ^2}\sqrt \rho  }}{{\partial {x^2}}}} \right) + \frac{{3{\epsilon^2}}}{\rho }\frac{\partial }{{\partial x}}\left( {\frac{{\rho \sqrt {1 + R_0^2{\rho ^{2/3}}} }}{{R_0^2{\rho ^{2/3}}}}} \right). \\
\end{array}
\end{equation}
where, $\beta=\alpha ({2^{5/3}}/5){(3{Z^2}/\pi )^{1/3}}$ and we have used $(1/{\rho}){\partial _x}{P_e(\rho)} = {\partial _x}\sqrt {1 + R^2}$ and $(1/{\rho}){\partial _x}{P_C(\rho)} = {\partial _x}(\beta R)$ and employed the following scalings in order to obtain the dimensionless equations ignoring the normalization bar notation
\begin{equation}\label{normal}
x \to \frac{{{V_{s}}}}{{{\omega _{pi}}}}\bar x,\hspace{3mm}t \to \frac{{\bar t}}{{{\omega _{pi}}}},\hspace{3mm}\rho \to \bar \rho{\rho_0},\hspace{3mm}V \to \bar V{V_{s}},
\end{equation}
where, $c$, ${\omega _{pi}} = \sqrt {{e^2}{n_{e0}}/(\varepsilon_0{m_i})}$ and ${V_{s}} = c\sqrt {{m_e}/{m_i}}$ are the vacuum light speed, the characteristic ion plasma frequency and ion sound-speed (this speed is much higher despite the name comparable to the Fermi-speed of an electron in a solid), respectively, and the parameter $\rho_{0}$ denotes the equilibrium plasma mass-density. We have also introduced new fractional plasma entities used frequently in forthcoming discission, namely, the quantum diffraction parameter, $H = \sqrt {{m_i}/2{m_e}} (\hbar {\omega _{pi}})/({m_e}{c^2})$, relativistic degeneracy parameter, $R_0=(\rho_0/\rho_c)^{1/3}$ \cite{akbari6}, and the normalized Zeeman energy, $\epsilon=\mu_B B_0/(m_e c^2)$.

\section{Coulomb Instability and Degeneracy Regimes}\label{extreme}

Now let us consider the special cases of $\frac{\epsilon}{R_0}\ll 1$ and $\frac{\epsilon}{R_0}\gg 1$ in the spin-contributed term in the momentum equation (Eq. (\ref{diff})). These limits will be shown to lead to distinct features in dispersion of linear spin-induced magnetosonic waves. Since, the normalized Zeeman energy parameter is solely defined by the external magnetic field strength, the two extreme degeneracy limits, namely, the nonrelativistic ($R_0\ll 1$) and relativistic ($R_0\gg 1$) degeneracy can be distinguished for a given magnetic field strength value, say, $B_0=10^{5}T$ ($\epsilon\simeq 10^{-5}$) typical of spin-dominated compact astrophysical objects. On the other hand, the reader should have already realized the elegant relationship between the quantum diffraction parameter, $H$, which is due to the electron tunneling and the relativistic degeneracy parameter, $R_0$, due to the increase in the electron number density, namely, $H = e\hbar \sqrt {{\rho_c}R_0^3/m_i\pi } /(2m_e^{3/2}{c^2})$. Fourier analyzing Eq. (\ref{diff}) results in the following dispersion relation for the linear magnetosonic waves in the coupled relativistically degenerate plasma case
\begin{equation}\label{disp}
\frac{{{\omega ^2}}}{{{k^2}}} = \frac{{{H^2}{k^2}}}{2} + \frac{{{R_0}}}{3}\left[ {\frac{{{R_0}}}{{\sqrt {1 + R_0^2} }} - \beta } \right] + \frac{{{\epsilon^2}}}{{{H^2}}}.
\end{equation}
The differences in dispersion relation (Eq. (\ref{disp})) between the two cases of $\frac{\epsilon}{R_0}\ll 1$ (relativistic degeneracy) and $\frac{\epsilon}{R_0}\gg 1$ (normal degeneracy) is readily apparent from the dispersion relation Eq. (\ref{disp}). It is observed that, for the case of relativistic degeneracy the last term in dispersion relation is negligibly small and the phase speed of the linear magnetosonic-wave is almost independent from Zeeman energy, $\epsilon$, while for the nonrelativistic plasma regime this is not the case. On the other hand, for the case of nonrelativistic degeneracy the first term in the dispersion relation is insignificant and the wave becomes dispersionless. The approximate dispersion regimes in extreme degeneracy limits can be summarized as below
\begin{equation}\label{limdisp}
\left\{ {\begin{array}{*{20}{c}}
{\frac{{{\omega ^2}}}{{{k^2}}} \approx \frac{{{H^2}{k^2}}}{2} + \frac{{{R_0}}}{3}\left[ {\frac{{{R_0}}}{{\sqrt {1 + R_0^2} }} - \beta } \right]} & {\left( {\frac{\epsilon}{{{R_0}}} \ll 1} \right)}  \\ {\frac{{{\omega ^2}}}{{{k^2}}} \approx \frac{{{\epsilon^2}}}{{{H^2}}} + \frac{{{R_0}}}{3}\left[ {\frac{{{R_0}}}{{\sqrt {1 + R_0^2} }} - \beta } \right]} & {\left( {\frac{\epsilon}{{{R_0}}} \gg 1} \right)}  \\
\end{array}} \right\}.
\end{equation}
It is apparent from Eq. (\ref{disp}) that the instability can set in when
\begin{equation}\label{ins}
\frac{{{H^2}{k^2}}}{2} + \frac{{{R_0}}}{3}\left[ {\frac{{{R_0}}}{{\sqrt {1 + R_0^2} }} - \beta } \right] + \frac{{{\epsilon^2}}}{{{H^2}}} < 0.
\end{equation}
This condition can be relaxed for the nonrelativistic ($R_0\ll 1$) long-wavelength ($k \ll 1$) limit to the following instability condition
\begin{equation}\label{ins}
\epsilon<\epsilon_m = H\sqrt {\frac{{{R_0}}}{3}\left[ {\beta  - \frac{{{R_0}}}{{\sqrt {1 + R_0^2} }}} \right]}.
\end{equation}
Thus, for every plasma composition (every value of $Z$) there exists a $R_0$-region in $R_0$-$\epsilon$ plane for a given Zeeman parameter $\epsilon<\epsilon_{m}$, where the instability sets-in for the magnetosonic linear waves and for $\epsilon>\epsilon_{m}$ the linear waves are completely stable. This effect has been discussed in detail in the next section, where, it has been shown that the width of instability region only loosely depends on the wavenumber, $k$ (e.g. see Fig. 4).

\section{Numerical Results an Analysis}\label{discussion}

It is clearly observed from Eq. (\ref{disp}) that, in the nonrelativistic degeneracy limit of our model the diffraction parameter, $H$, is negligibly small, hence, the magnetosonic waves are almost dispersionless, while, for the relativistic degeneracy regime the values of, $H$ are relatively large and the linear waves become dispersive. The dispersion in this regime becomes more pronounced for higher values of wave number, $k$. In Fig. 1, the variation of paramagnetic susceptibility $\chi_p$, given by Eq. (\ref{QP}), versus the normalized plasma mass-density, $\rho_0/\rho_c$ is shown. It is observed that the increase in the mass-density (equivalently increase in the relativistic degeneracy parameter of the plasma) causes the increase in the relativistic susceptibility value.  Figure 2 shows the variation of the magnetosonic wave dispersion relation due to variations in the relativistic degeneracy parameter, $R_0$ (presenting the normalized plasma mass-density) and the normalized Zeeman energy, $\epsilon$ (presenting the magnetic field-strength) in the two distinct nonrelativistic and relativistic degeneracy limits in the absence of Coulomb interactions ($\beta=0$). From Fig. 2(a) it is apparent that in the nonrelativistic regime the speed of linear magnetosonic waves decrease with increase in the relativistic degeneracy parameter for a given fixed value of the magnetic field parameter, $\epsilon$.

On the other hand, it is observed from Fig. 2(b) that (in the absence of Coulomb interaction) increase in the strength of the magnetic field leads to the increase in the wave speed in this regime. However, as it is revealed from Fig. 2(c), in the relativistic degeneracy regime, the increase in the normalized plasma mass-density (i.e. increase in the relativistic degeneracy parameter, $R_0$) results in an increase in the linear magnetosonic speed which is opposite to the case of nonrelativistic degeneracy case shown in Fig. 2(a). Furthermore, increase in the magnetic field strength in relativistic degeneracy regime in ineffective in changing the linear wave-speed for a fixed normalized $R_0$-parameter which is contrary to the case of nonrelativistic degeneracy (e.g. compare Figs. 2(b) and 2(d)).

Distinguished feature is observed in Fig. 2(e) where a minimum in phase speed exists at a given value of relativistic degeneracy, $R_{0m}$. This minimum ($R_{0m}$) occurs in nonrelativistic degeneracy regime value of which depends strongly on the magnitude of the Zeeman energy parameter, $\epsilon$. The occurrence of the minimum in wave speed is related to the competition of the positive electron degeneracy pressure and the negative pressure-like spin paramagnetism effect. As it is revealed comparing Figs. 2(d) and 2(e) the value of $R_{0m}$ increases with increase in the strength of the magnetic field. The corresponding value of $R_{0m}\simeq 0.1$ ($n_0\simeq 5.9\times10^{26} cm^{-3}$) is obtained for $\epsilon=10^{-5}$ ($B_0\simeq10^5 T$), while, this value increases to $R_{0m}\simeq 0.25$ ($n_0\simeq 9.2\times10^{27} cm^{-3}$) for $\epsilon=10^{-4}$ ($B_0\simeq10^6 T$). Figure 3(a), clearly confirms the above statement. On the other hand, Fig. 3(b) reveals that the value of the minimum phase-speed of the linear waves is nearly independent of the value of the wave-number, $k$. The observed differences in dispersion of linear magnetosonic waves in different plasma degeneracy regimes mark distinct behavior of linear magnetosonic wave in dense paramagnetic quantum plasmas which is typical of the astrophysical situations and can improve our understanding of the linear wave propagation mechanism in superdense highly magnetized collapsing white dwarfs or neutron stars. Such distinctive features has already been suggested to be present in nonlinear wave dynamics of ion-acoustic solitons in Thomas-Fermi plasmas \cite{akbari7, akbari8}.

Figure 4(a) shows and compares the variation of dispersion relations for $^4He$ and $^{56}Fe$ plasma compositions for a fixed value of Zeeman energy parameter. It is revealed that for the $^{56}Fe$ magnetized plasma there is a instability gap for a range of relativistic degeneracy parameter, $R_0$ for which magnetosonic linear waves can not propagate. The width of the instability gap seems not to vary significantly with the wavenumber value, $k$. Figure 4(b) reveals that for the $^{56}Fe$ composition also the instability sets-in when the magnetic field strength is lowered. Remarkably it is revealed, comparing Figs. 4(a) and 4(b) that the change in the magnetic field strength for a fixed plasma composition only affects the nonrelativistic degeneracy part of the plasma, while, the composition change in plasma for fixed magnetic field strength impacts the relativistic degeneracy part. The instability region for compositions $Z=2,4,6,26$ are shown in Fig. 5. It is also remarked that the instability area increases with increase in the atomic-number of plasma composition. As it is evident from Fig. 5, for every composition there is a minimum Zeeman parameter, $\epsilon$, below which the instability of linear magnetosonic waves occur. For a given $\epsilon$-value below this minimum the instability is bounded by two values of degeneracy parameter $R_0$. The $R_0$ range for instability increases as the magnetic field strength is lowered, however, for iron plasma composition the situation is different and the plasma becomes linearly unstable for all magnetic field strength values above a minimum $R_{0m}$ value. These features might be crucial on the stability of compact stars such as white dwarfs and neutron star crusts. For instance, the crust of a neutron star or a white-dwarf core is believed to be iron-rich while they are also relativistically degenerate and might be highly magnetized. Also, the outer regions of a white-dwarf star may contain degenerate carbon or helium. There has been many compact star reports with strong internal or external magnetic fields \cite{crut, kemp, put, jor}.

\newpage

\textbf{FIGURE CAPTIONS}

\bigskip

Figure-1

\bigskip

The paramagnetic susceptibility, $\chi_p$, versus normalized electron number-density, $n_e/N_0$, for a Fermi-Dirac plasma.

\bigskip

Figure-2

\bigskip

(Color online) Variation of the magnetoacoustic dispersion curves with respect to the changes in the relativistic degeneracy parameter, $R_0$ and the normalized Zeeman-energy, $\epsilon$, in the nonrelativistic plasma degeneracy regime (Figs. 1(a) and 1(b)) and in the relativistic plasma degeneracy regime (Figs. 1(c) and 1(d)). The minimum dispersion slope (wave speed) is shown for two different values of magnetic field strength (Figs. 1(e), 1(f)).

\bigskip

Figure-3

\bigskip

(Color online) Variation of the magnetoacoustic wave phase speed with respect to the relativistic degeneracy parameter, $R_0$ for differnt values of the normalized Zeeman-energy, $\epsilon$ (a), the wave-number, $k$ (b). The dash size of curve increases appropriately as the varied parameter in each plot is increased.

\bigskip

Figure-4

\bigskip

(Color online) Figure 4(a) shows the complete stability for $He^4$ (top surface) but partial stability for $^{56}Fe$ (bottom surfaces) plasma composition at $\epsilon=10^{-7}$. Figure 4(b) shows the instability inset for $^{56}Fe$ as the Zeeman parameter (magnetic-field strength) decreases from $\epsilon=10^{-6}$ to $\epsilon=10^{-7}$.

\bigskip

Figure-5

\bigskip

(Color online) Unstable linear magnetosonic wave regions in $R_0$-$\epsilon$ plane for different plasma compositions, $Z=2,4,6,26$ with condition $R_0<<1$. The dashed areas (respectively from small to large correspond to $Z=2,4,6,26$) indicate the unstable regions for different plasma compositions.

\bigskip

\end{document}